# Origin of the large differences in high-pressure stability and superconductivity between ThH$_9$ and ThH$_{18}$

Shichang Yao[†,‡], Chongze Wang[‡], Hyunsoo Jeon[‡], Liangliang Liu[†,§], Jin Mo Bok[ℵ], Yunkyu Bang[ℵ,ℵ], Yu Jia[†,§], and Jun-Hyung Cho[*,‡]

[†]Joint Center for Theoretical Physics, School of Physics and Electronics, Henan University, Kaifeng 475004, People's Republic of China

[‡]Department of Physics and Research Institute for Natural Science, Hanyang University, 222 Wangsimni-ro, Seongdong-Ku, Seoul 04763, Republic of Korea.

[§]Key Laboratory for Special Functional Materials of the Ministry of Education, Henan University, Kaifeng 475004, People's Republic of China

[ℵ]Department of Physics, Pohang University of Science and Technology, Pohang 37673, Republic of Korea.

[ℵ]Asia Pacific Center for Theoretical Physics (APCTP), Pohang-si, Gyeongsangbuk-do 37673, Republic of Korea

**ABSTRACT:** Recently, the thorium hydride ThH$_9$ possessing an H-rich clathrate structure has been experimentally synthesized to exhibit a superconducting transition temperature $T_c$ of 146 K at 170−175 GPa, while the more H-rich clathrate thorium hydride ThH$_{18}$ was theoretically predicted to reach a $T_c$ of 296 K at 400 GPa. Using first-principles calculations, we find that ThH$_9$ has a more ionic character between Th atoms and H cages than ThH$_{18}$ and that the latter has a more substantial hybridization of the Th $6p$ semicore and H $1s$ states than the former. These different bonding characteristics of ThH$_9$ and ThH$_{18}$ are associated with their stability at very different pressures. Furthermore, we reveal that (i) the H-derived density of states at the Fermi level $E_F$ is about two times larger in ThH$_{18}$ than in ThH$_9$, (ii) the average squared phonon frequency of H atoms is ≈29% higher in ThH$_{18}$ than in ThH$_9$, and (iii) the Fermi-surface average squared electron-phonon matrix element is similar between the two hydrides. Consequently, the electron-phonon coupling constant of ThH$_{18}$ becomes much greater than that of ThH$_9$, leading to a significant $T_c$ difference between the two thorium hydrides. Our findings not only provide an explanation for the very large differences in the stabilization pressure and superconducting transition temperature between ThH$_9$ and ThH$_{18}$ but also have important implications for the design of H-rich, high-$T_c$ clathrate metal hydrides.

The realization of room-temperature superconductivity (SC) is one of the grand challenges in condensed matter physics[1–3]. An early theoretical proposal[4] of high-temperature SC began with the atomic metallic hydrogen that can be transformed from molecular insulating hydrogen under high pressures[5,6]. However, the experimental realization of atomic metallic hydrogen has been very challenging[1–3], because it requires too high pressures over ≈500 GPa, which is an upper limit attainable with diamond anvil cells[7,8]. To achieve such a hydrogen-driven metallic state at relatively lower pressures, an alternative approach has been devised using hydride materials[9] in which hydrogen atoms can be "chemically precompressed" through interactions with other constituent atoms. This approach has recently been demonstrated to be viable for the realization of near-room temperature SC at much lower pressures below 200 GPa[10–15].

Motivated by the precedent theoretical predictions of high superconducting transition temperature $T_c$ in compressed hydrides[16–26], intensive experimental efforts have been made for the synthesis of various hydrides such as sulfur hydride H$_3$S[10], rare-earth/actinide hydrides MH$_n$ (M = La[11,12], Y[13–15], Ce[27–29], and Th[30]), and alkaline earth hydride CaH$_6$[31,32]. Specifically, the latter two families of metal hydrides have H-rich clathrate structures, where each metal atom is surrounded by the H cage composed of a large number of H atoms, i.e., 24, 29, and 32 H atoms for n = 6, 9, and 10, respectively. These H-rich clathrate metal hydrides exhibit a wide range of $T_c$ depending on metal elements. For example, LaH$_{10}$ was experimentally observed to exhibit a $T_c$ of 250−260 K at pressures of 170−190 GPa[11,12] and subsequently, YH$_6$ was measured to exhibit $T_c$ = 220 K at 166 GPa[13,14], YH$_9$, $T_c$ = 243 K at 201 GPa[14,15], ThH$_{10}$, $T_c$ = 159 K at 174 GPa[30], and CaH$_6$, $T_c$ = 215 K at 170 GPa[31,32].

Recently, Zhong et al.[33] used a crystal structure search method to find a new class of extremely H-rich clathrate rare earth/actinide hydrides MH$_{18}$ (M = Y, La, Ce, Ac, and Th). Among them, the heaviest element Th hydride ThH$_{18}$ was predicted to exhibit a $T_c$ of 296 K at 400 GPa. Interestingly, these values of $T_c$ and stabilization pressure in ThH$_{18}$ are much higher than the experimental data ($T_c$ = 146 K at 170 GPa) in ThH$_9$ having half of hydrogen content[30]. Zhong et al.[33] explained the existence of the higher $T_c$ in ThH$_{18}$ in terms of the increased electronic density of states (DOS) at the Fermi level $E_F$, the large phonon energy scale of the vibration modes, and the resulting enhanced electron-phonon coupling (EPC). Although these factors are essential for increasing $T_c$ in conventional phonon-mediated BCS[34] superconductors, more quantitative analysis is desired to explain the significant $T_c$ difference between ThH$_9$ and ThH$_{18}$, as discussed below. Furthermore, understanding the underlying mechanism of a significant difference in the stabilization pressure between ThH$_9$ and ThH$_{18}$ is valuable in searching for H-rich, high-$T_c$ clathrate metal hydrides.

In this paper, based on first-principles calculations, we identify the origin of the very large differences in the stabilization pressure and $T_c$ between ThH$_9$ and ThH$_{18}$. For this, we compare the structural, electronic, phononic, and superconducting properties of ThH$_9$ and ThH$_{18}$ compressed at 130 and 400 GPa, respectively. Compared to ThH$_{18}$, ThH$_9$ has a greater ionic character with more charge transfer from Th to H atoms, which results in stabilization at relatively lower pressures due to increased chemical precompression. Meanwhile, ThH$_{18}$ has substantially delocalized Th $6p$ semicore states at a high pressure of 400 GPa, leading to their strong hybridization with the H $1s$ state. These different bonding characteristics of ThH$_9$ and ThH$_{18}$ reflect a significant difference in their stabilization pressures. Furthermore, we reveal that ThH$_{18}$ has about two times larger H-derived DOS at $E_F$ and ≈29% higher average squared H-derived phonon frequency than ThH$_9$, but the two hydrides have similar Fermi-surface average squared electron-phonon matrix elements. As a result, the EPC constant λ and $T_c$ of ThH$_{18}$ are estimated as 2.08 and 269 K, respectively, which are greater than those (1.55 and 151 K) of ThH$_9$.

Our DFT calculations were performed using the Vienna ab initio simulation package with the projector-augmented wave method[35–37]. For the exchange-correlation energy functional, we employed the generalized-gradient approximation of Perdew-Burke-Ernzerhof[38]. We treated Th $6s^26p^65f^16d^17s^2$ and H $1s^1$ as valence electrons, including $6s^26p^6$ semicore electrons for Th. A plane-wave basis was used with a kinetic energy cutoff of 500 eV. The $k$-space integration of ThH$_9$ (ThH$_{18}$) was done with the 18×18×12 (8×8×8) $k$ points for the structure optimization and the 24×24×18 (18×18×18) $k$ points for the DOS calculation. All atoms were allowed to relax along the calculated forces until all the residual force components were less than 0.001 eV/Å. We calculated the phonon spectrum of ThH$_9$ (ThH$_{18}$) using the QUANTUM ESPRESSO package[39], where ultrasoft pseudopotentials[40], a kinetic energy cutoff of 1088 eV, 18×18×12 (8×8×8) $k$ points, and 6×6×6 (2×2×2) $q$ points were employed. For the EPC calculation, we used the software EPW[41,42] with the 12×12×8 (6×6×6) $q$ and 24×24×16 (24×24×24) $k$ points for ThH$_9$ (ThH$_{18}$).

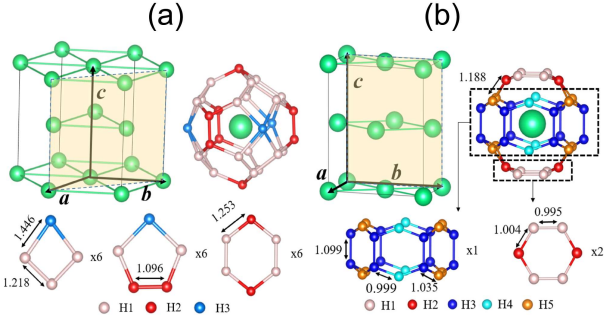

FIG. 1. Optimized structures of (a) ThH$_9$ at 130 GPa and (b) ThH$_{18}$ at 400 GPa. The Th sublattice in ThH$_9$ (ThH$_{18}$) forms the hcp (centered rectangular) lattice with the H$_{29}$ (H$_{36}$) cage surrounding each Th atom. The H$_{29}$ cage is composed of six tetragon, six pentagon, and six hexagon rings, while the H$_{36}$ cage is composed of a strip of six hexagon rings and two wrinkled hexagon rings. There are three (five) different types of H atoms in ThH$_9$ (ThH$_{18}$). The (110) and (100) planes are drawn in the hcp and centered rectangular lattices, respectively.

We begin by optimizing the structures of ThH$_9$[30] and ThH$_{18}$[33] using first-principles DFT calculations. Figures 1(a) and 1(b) show the optimized structures of ThH$_9$ and ThH$_{18}$ at 130 GPa and 400 GPa, respectively. Hereafter, we focus on the comparison of the structural, electronic, phononic, and superconducting properties of these two compressed structures. For ThH$_9$, the Th sublattice forms the hcp lattice with the lattice constants $a = b = 3.726$ Å and $c = 5.565$ Å [see Fig. 1(a)], where the H$_{29}$ clathrate cage surrounding a Th atom consists of six tetragon, six pentagon, and six hexagon rings. Meanwhile, the Th sublattice in ThH$_{18}$ forms the centered rectangular lattice with $a = 3.357$ Å, $b = 5.826$ Å and $c = 7.180$ Å [see Fig. 1(b)], where the H$_{36}$ clathrate cage surrounding a Th atom consists of the strip of six hexagon rings and two wrinkled hexagon rings above and below the strip. It is worth noting that ThH$_9$ has four different H−H bond lengths of 1.096, 1.218, 1.253, and 1.446 Å [see Fig. 1(a)]. By contrast, the H−H bond lengths in ThH$_{18}$ are relatively shortened with a range of 0.995−1.188 Å [see Fig. 1(b)], which in turn raises the maximum H-derived phonon frequency compared to that of ThH$_9$, as will be shown later.

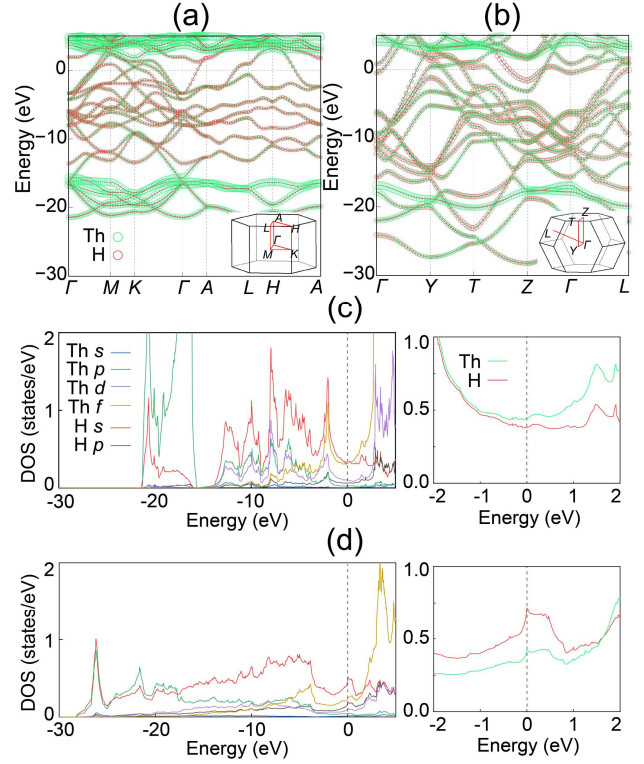

FIG. 2. Calculated band structures of (a) ThH$_9$ and (b) ThH$_{18}$. The projected bands onto Th and H atoms are represented by circles whose radii are proportional to the weights of the corresponding atoms. The energy zero represents $E_F$. The Brillouin zones corresponding to the primitive cells of ThH$_9$ and ThH$_{18}$ are included in (a) and (b), respectively. The PDOS results for the Th and H orbitals of ThH$_9$ and ThH$_{18}$ are displayed in (c) and (d), respectively. The right panels in (c) and (d) enlarge the PDOS sums for Th and H atoms around $E_F$.

Figures 2(a) and 2(b) show the calculated band structures of ThH$_9$ and ThH$_{18}$, respectively. It is seen that the band projections onto Th and H atoms in ThH$_9$ (ThH$_{18}$) represent their strong hybridization over the entire energy range between −21.4 (−28.2) eV and $E_F$. To explore the Th-H hybridization in detail, we present the partial DOS (PDOS) of ThH$_9$ and ThH$_{18}$ in Figs. 2(c) and 2(d), respectively. In contrast to ThH$_9$ having the rather localized Th $6p$ semicore states around −20 eV below $E_F$, ThH$_{18}$ has the substantially delocalized Th $6p$ semicore states that strongly hybridize with the H $1s$ state in a wide energy range from ≈−28 to ≈−17 eV. Such a large delocalization of Th $6p$ semicore electrons in ThH$_{18}$ is likely due to the increased interactions between Th atom and its surrounding H atoms at a high pressure of 400 GPa. Note that the distances between Th and H atoms in ThH$_{18}$ are shortened with a range of 1.88−2.04 Å, compared to those (2.08−2.23 Å) in ThH$_9$. Therefore, the strong hybridization of the Th semicore and H 1$s$ states in ThH$_{18}$ could be an essential ingredient for the stabilization of the large H$_{36}$ clathrate cage. The right panels in Figs. 2(c) and 2(d) display a closeup of the Th- and H-derived DOS around $E_F$. For ThH$_{18}$, we find a van Hove singularity at $E_F$, giving rise to an H-derived DOS of 0.715 states/eV at $E_F$. This magnitude of H-derived DOS is ≈1.9 times larger than 0.382 states/eV for ThH$_9$. Such a large difference in the H-derived DOS at $E_F$ between ThH$_9$ and ThH$_{18}$ mainly contributes to a significant difference in their $T_c$ values, as discussed below.

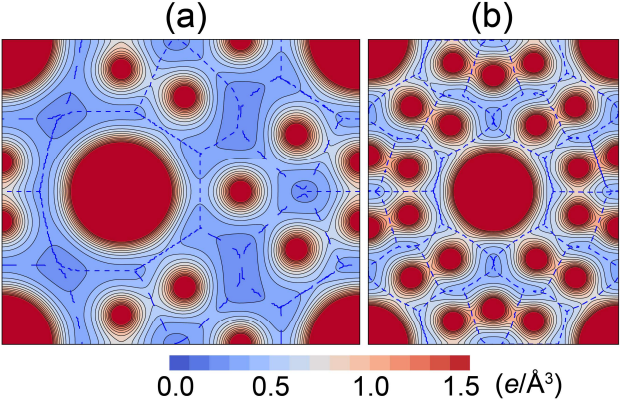

FIG. 3. Calculated total charge densities of (a) ThH$_9$ and (b) ThH$_{18}$, together with the Bader basins of Th and H atoms. The charge densities in (a) and (b) are plotted on the (110) and (100) planes [see Figs. 1(a) and 1(b)] with the contour spacing of 0.1 $e$/Å$^3$, respectively.

It is well established that H-rich clathrate metal hydrides have an ionic character between metal and H atoms due to their charge transfer[43,44]. To estimate the charge transfer from Th to H atoms, we analyze their Bader charges[45] in ThH$_9$ and ThH$_{18}$. Figures 3(a) and 3(b) show the calculated total charge densities of ThH$_9$ and ThH$_{18}$ with the Th/H Bader basins[45], respectively. Here, the Bader basins are obtained from a computation of the gradient of the total charge density[46]. The estimated cationic/anionic charges within the Th/H Bader basins are summarized in Table 1. We find that the Th cationic charge in ThH$_9$ is 1.676$e$, larger than that (1.437$e$) in ThH$_{18}$. Accordingly, the sum of the anionic charges of different H atoms in ThH$_9$ is larger in magnitude than that in ThH$_{18}$. Therefore, ThH$_9$ has a more ionic character between Th and H atoms, compared to ThH$_{18}$. The stronger Th−H ionic bond in ThH$_9$ gives rise to an enhancement of chemical precompresssion, which in turn lowers its stabilization pressure compared to ThH$_{18}$. Meanwhile, ThH$_{18}$ having a relatively weakened Th−H ionic bond likely stabilizes the extremely H-rich H$_{36}$ cage via the above-mentioned strong hybridization between the Th semicore and H 1$s$ states at a high pressure of 400 GPa.

**Table 1. Estimated cationic/anionic charges (in the unit of $e$) within the Bader basins of Th and H [see Figs. 1(a) and 1(b)] atoms in ThH$_9$ and ThH$_{18}$.**

|  | Th | H1 | H2 | H3 | H4 | H5 |
|---|---|---|---|---|---|---|
| ThH$_9$ | 1.676 | -0.264(1) | -0.091(2) | -0.205(6) | | |
| ThH$_{18}$ | 1.437 | -0.098(4) | -0.019(4) | -0.022(4) | -0.147(4) | -0.146(2) |

The numbers in parentheses represent the number of different H atoms in each formula unit.

Next, we examine the phonon spectra of ThH$_9$ and ThH$_{18}$ using density functional perturbation theory calculations[39]. Figures 4(a) and 4(b) display the calculated phonon dispersions of ThH$_9$ and ThH$_{18}$, respectively. The projected DOS onto Th and H atoms shows that for both ThH$_9$ and ThH$_{18}$, the acoustic phonon modes of Th atoms are well separated from the optical phonon modes of H atoms. We find that the H-derived phonon modes in ThH$_9$ are distributed between 496 and 1655 cm$^{-1}$ [see Fig. 4(a)], while those in ThH$_{18}$ reside in a wide frequency range from 397 to 2587 cm$^{-1}$ [see Fig. 4(b)]. Therefore, the H-derived phonon modes in ThH$_{18}$ exhibit not only a hardening of the high-frequency regime but also a softening of the low-frequency regime, compared to those in ThH$_9$. The former (latter) hardening (softening) is likely caused by the shorter H−H (Th−H) bond lengths in ThH$_{18}$ [see Fig. 1(b)]. As a result, we find that the logarithmic average $\omega_{log}$ of H-derived phonon frequencies in ThH$_{18}$ is 1034 cm$^{-1}$, higher than that (885 cm$^{-1}$) in ThH$_9$ (see Table 2).

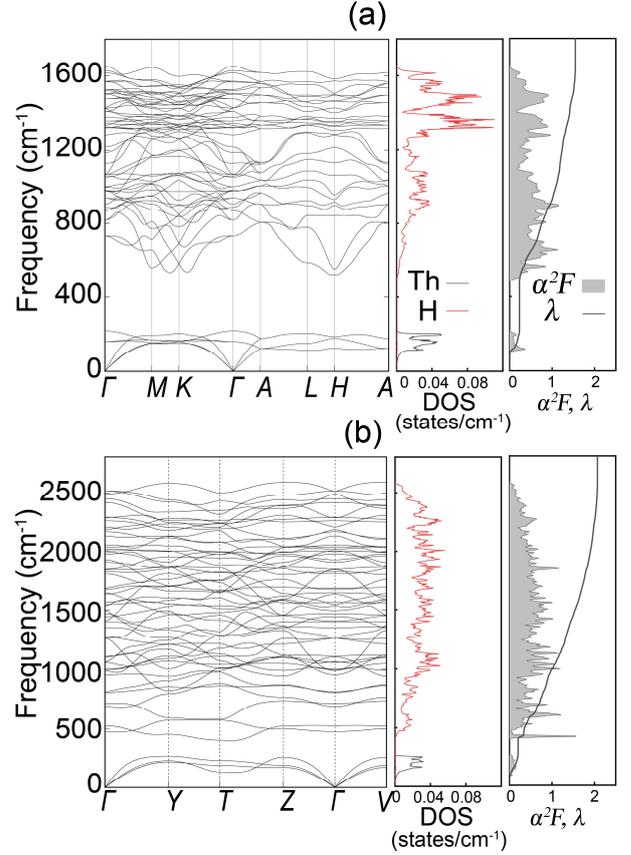

FIG. 4. Calculated phonon spectra, phonon DOS projected onto Th and H atoms, $\alpha^2F(\omega)$, and $\lambda(\omega)$ of (a) ThH$_9$ and (b) ThH$_{18}$.

Using the isotropic Migdal-Eliashberg formalism[48–50], we calculate the Eliashberg spectral function $\alpha^2F(\omega)$ and integrated EPC constant $\lambda(\omega)$ as a function of phonon frequency. Figures 4(a) and 4(b) show the comparison of $\alpha^2F(\omega)$ and $\lambda(\omega)$ between ThH$_9$ and ThH$_{18}$. We find that for the two hydrides, all the phonon modes including the Th-derived acoustic and H-derived optical modes contribute to increasing $\lambda(\omega)$. For ThH$_9$ (ThH$_{18}$), the Th-derived acoustic modes are estimated to contribute to ≈15 (10)% of the total EPC constant $\lambda = \lambda(\infty)$, while the H-derived optical modes contribute to ≈85 (90)% of $\lambda$. Specifically, $\lambda(\omega)$ arising from H-derived phonon modes monotonically increases with increasing frequency. Therefore, ThH$_{18}$ having higher H content reaches a much larger value of $\lambda$ = 2.08, compared to that (1.55) of ThH$_9$ [see Figs. 4(a) and 4(b)]. By numerically solving the isotropic Eliashberg equations[49] with the typical Coulomb pseudopotential parameter of $\mu^* = 0.1$[33], we obtain the temperature dependence of superconducting gap $\Delta$ (see Fig. 5). For ThH$_9$, $\Delta$ closes at a $T_c$ of $\approx$151 K at 130 GPa, close to the measured[30] value of 146 K at 170−175 GPa. Meanwhile, for ThH$_{18}$, the estimated $T_c$ is as high as $\approx$269 K, being well comparable with that (296 K) of a previous theoretical calculation[33].

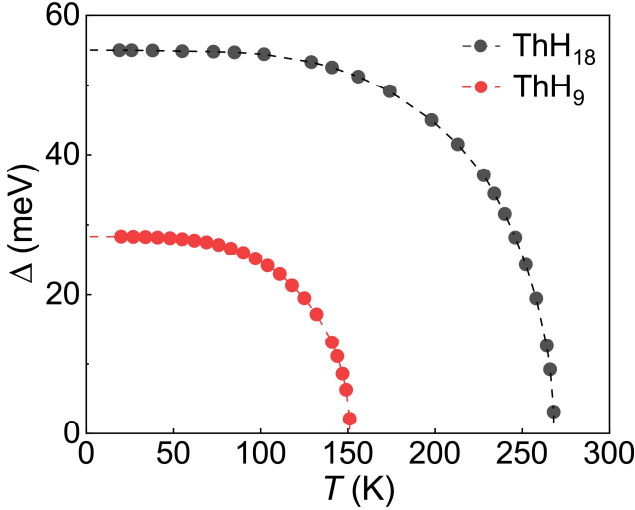

FIG. 5. Calculated superconducting gaps $\Delta$ of ThH$_9$ and ThH$_{18}$ as a function of temperature with $\mu^* = 0.1$.

To understand why there is a large difference in $\lambda$ between ThH$_9$ and ThH$_{18}$, we compare the contributions of the underlying components that determine $\lambda$. Since the frequency ranges of the Th-derived acoustic and H-derived optical modes are well separated [Figs. 4(a) and 4(b)], $\lambda$ can be expressed as a sum of its atom-specific components $\lambda_j$ ($j$ = Th, H) according to the McMillan-Hopfield theory[50–52]: i.e.,

$$\lambda = \int \frac{2}{w} \alpha^2 F(w)\, dw \;\to\; \sum_j \frac{N_j(E_F) I_j^2}{M_j \omega_{2,j}^2}, \qquad (1)$$

where $N_j(E_F)$ is the Th- or H-derived DOS at $E_F$, $I_j^2$ is the atom-specific Fermi-surface average squared electron-phonon matrix element, $M$ is the atomic mass, and $\omega_{2,j}^2$ is the atom specific average squared phonon frequency. Therefore, $\lambda_j$ is composed of the electronic part $N_j(E_F) I_j^2$ in the numerator of Eq. (1) and the phonon part $M_j \omega_{2,j}^2$ in the denominator. For ThH$_9$ and ThH$_{18}$, each component contributing to $\lambda_j$ is listed in Table 2. Using the McMillan-Allen-Dynes formula[53], we can estimate $T_c$ from $\lambda_j$ and $\omega_{\log}$ associated with each atom[52]. Although the McMillan-Allen-Dynes formula usually underestimates $T_c$[54–58], its estimation may provide a qualitative aspect of how largely certain phonon modes contribute to $T_c$. As shown in Table 2, the acoustic phonon modes of Th atoms contribute to an increase in $\lambda$, but they hardly increase $T_c$, similar to the cases of other high-$T_c$ hydrides[52, 56]. Since $\lambda_{Th}$ is insensitive to an increase in $T_c$, we examine how $N_H(E_F)$, $I_H^2$, and $M_H \omega_{2,H}^2$ for H atom contribute to $\lambda_H$ that mostly determines $T_c$. In Table 2, we find that (i) $N_H(E_F)$ is $\approx$1.9 times larger in ThH$_{18}$ than in ThH$_9$, (ii) $M_H \omega_{2,H}^2$ is $\approx$29% larger in ThH$_{18}$ than in ThH$_9$, and (iii) $I_H^2$ in ThH$_{18}$ is similar between the two hydrides. Consequently, we can say that for ThH$_{18}$, $\lambda_H$ is dominantly increased by $N_H(E_F)$ despite its lowering via $M_H \omega_{2,H}^2$ [see Eq. (1)], resulting in a higher $T_c$ compared to ThH$_9$.

**Table 2.** Estimated components $N_j(E_F)$, $I_j^2$, and $M_j \omega_{2,j}^2$ that determine $\lambda_j$ ($j$ = Th, H) in ThH$_9$ and ThH$_{18}$.

| | | $\omega_{\log}$ (cm$^{-1}$) | $N(E_F)$ (eV$^{-1}$) | $I^2$ (eV$^2$Å$^{-2}$) | $M\omega_2^2$ (eVÅ$^{-2}$) | $\lambda$ | $T_c$ (K) |
|---|---|---|---|---|---|---|---|
| ThH$_9$ | Th | 135 | 0.444 | 7.282 | 13.876 | 0.233 | 0 |
| | H | 885 | 0.382 | 1.461 | 0.423 | 1.319 | 139 |
| ThH$_{18}$ | Th | 151 | 0.421 | 7.904 | 16.153 | 0.206 | 0 |
| | H | 1034 | 0.715 | 1.428 | 0.545 | 1.874 | 240 |

The estimated values for $\omega_{\log}$ and $T_c$ arising from Th and H atoms are also given. Here, $T_c$ values are estimated using the McMillan-Allen-Dynes formula[53].

In summary, using first-principles calculations, we have investigated the structural, electronic, phononic, and superconducting properties of ThH$_9$ and ThH$_{18}$ which showed very large differences in the stabilization pressure and $T_c$[30,33]. We found that ThH$_9$ has a greater ionic bonding character due to more charge transfer from Th to H atoms, thereby inducing a relatively larger chemical precompression. Meanwhile, ThH$_{18}$ has a more substantial hybridization of the Th 6$p$ semicore and H 1$s$ states, which is likely associated with the stabilization of the extremely H-rich H$_{36}$ clathrate cage at higher pressures. Furthermore, we revealed that ThH$_{18}$ has about two times larger H-derived DOS at $E_F$ than ThH$_9$, which dominantly contributes to enhancing $\lambda$ and $T_c$. However, the electron-phonon matrix element was found to make minor contributions to the change of $\lambda$ between ThH$_9$ and ThH$_{18}$. We thus demonstrated that the difference in the bonding character between Th atom and its surrounding H cage reflects the different stabilization pressures of ThH$_9$ and ThH$_{18}$ and that the H-derived DOS at $E_F$ plays a crucial role in the very large difference in $T_c$ between the two Th hydrides.

## ASSOCIATED CONTENT

## AUTHOR INFORMATION

**Corresponding Author**

*chojh@hanyang.ac.kr

**Author Contributions**

S. Y. and C. W. contributed equally to this work.

**Notes**

The authors declare no competing financial interests.

## ACKNOWLEDGMENT

This work was supported by the National Research Foundation of Korea (NRF) grant funded by the Korean Government (Grant Numbers 2022R1A2C1005456), by BrainLink program funded by the Ministry of Science and ICT through the National Research Foundation of Korea (2022H1D3A3A01077468), and by the National Natural Science Foundation of China (Grant No. 12074099). The calculations were performed by the KISTI Supercomputing Center through the Strategic Support Program (Program No. KSC-2022-CRE-0073) for the supercomputing application research.